\documentclass{llncs}
\usepackage{makeidx}  
\usepackage{graphicx}
\usepackage{subfigure}
\usepackage{amsmath}

\begin{document}
\frontmatter          
\pagestyle{headings}  
\addtocmark{Hamiltonian Mechanics} 
\mainmatter              
\title{Real-time Deformation of Soft Tissue Internal Structure with Surface Profile Variations using Particle System}

\titlerunning{RealtimeSoftTissueDeformation}  
%
\author{Haoyin Zhou\inst{1} \and Eva Gombos\inst{1} \and Mehra Golshan\inst{1} \and Jayender Jagadeesan\inst{1}}
%
%
%
\institute{Brigham and Women's Hospital, Harvard Medical School
}

\maketitle              

\begin{abstract}

Intraoperative observation of tissue internal structure is often difficult. Hence, real-time soft tissue deformation is essential for the localization of tumor and other internal structures. We propose a method to simulate the internal structural deformations in a soft tissue with surface profile variations. The deformation simulation utilizes virtual physical particles that receive interaction forces from the surface and other particles and adjust their positions accordingly. The proposed method involves two stages. In the initialization stage, the three-dimensional internal structure of the surface mesh is uniformly sampled using the particle expansion and attracting-repelling force models whilst simultaneously building the internal particle connections. In the simulation stage, under surface profile variations, we simulate the internal structural deformation based on a deformation force model that uses the internal particle connections. The main advantage of this method is that it greatly reduces the computational burden as it only involves simplified calculations and also does not require generating three-dimensional meshes.  Preliminary experimental results show that the proposed method can handle up to 10,000 particles in 0.3s.

\end{abstract}

\section{Introduction}

In the operating room, real-time intraoperative localization of the tumor and other internal structures of a tissue is essential and challenging. Compared to the surgical preparation and planning stage, when the structure of a diseased tissue can be accurately obtained by the combination of many diagnostics technologies, the intraoperative localization of internal structures is more difficult. For example, only 50\% of non-palpable tumors are visible by intraoperative ultrasound in the breast \cite{Pleijhuis}. In contrast, in the process of many surgeries, the tissue surface can be directly observed and be modeled by construction methods. For example, Conley et al used a laser scanner to reconstruct the skin surface of the breast \cite{Conley}. Hence, it is important to obtain the internal structural deformations in a soft tissue according to surface profile variations.

Deformation simulation is a key problem in the computer graphics field and many approaches have been proposed \cite{Le}\cite{Chao}. Among these methods, particle system-based methods have been proven to be efficient and are widely used \cite{Macklin}\cite{MüllerSolid}. The main idea of particle system is to utilize virtual physical particles which interact with each other and move accordingly. Hence, particle system has an advantage of being more close to the reality.

For performing particle system-based calculation, it is a prerequisite to know which particles have direct connections between them. Many deformation methods use polygon meshes to represent the connections \cite{Diziol}. However, for a three-dimensional (3D) structure, the Delaunay-based tetrahedral meshes generation is quite time consuming \cite{Shewchuk}\cite{SiTetGen}. Besides, the Delaunay-based methods need vertices as a prerequisite for performing mesh generation. Another type of deformation methods defines the connections by Euclidean distance thresholding \cite{MüllerMeshless}, which greatly reduced the computation burden. For both types of deformation methods, the initial uniformly distributed particles with connections would improve the accuracy.

With the built connections, the mass spring damper system \cite{EBERLY} is often employed for calculating the interaction forces between particles. However, the mass spring damper system has a significant drawback of not being able to take into account the variation in rotation angles directly. Although this drawback can be compensated by the implicit triangular stability in most cases, the mass spring damper system cannot handle situations with a large deformation.

In this paper, we employ this concept of particle system and utilize particles to represent physical units of a tissue. The main contribution of this paper are twofold: first, a novel sampling algorithm is proposed to construct virtual physical internal connections of a tissue with uniformly distributed particles, second; a novel local deformation description algorithm is proposed for generating deformation forces between particles. Both algorithms have been shown to be fast and efficient.


\section{Uniformly Sampling}

We want to down-sample the object internal structure with a given number of particles for reducing the computational burden of the subsequent deformation estimation. For this purpose, we choose uniformly sampling which can make the deformation simulation smooth. Further, because in a real world situation, physical particles only have direct interactions when they come in contact, and deformations are propagated through connections defined by this contacting relationships. Hence, it is necessary to build the virtual physical connections between particles for the deformation simulation.

Our uniformly sampling algorithm utilizes the conception of particle system, where particles receive forces from the object surface and their neighboring particles and move accordingly. For particle $p_i$, the relationship between its motion and the received forces is given as follows:

\begin{equation} \label{eqmotionmodel}
{\rm{d}}{p_i} = {{\bf{F}}_i}/m
\end{equation}

\begin{equation} \label{eqFisurfaceandParticle}
{\bf{F}}_i = \mu {\bf{F}}_{surface \rightarrow i} + \sum\limits_{j \in {\Omega _i}} {{{\bf{F}}_{j \rightarrow i}}}
\end{equation}

\noindent where ${\bf{F}}_i$ is sum of all forces it receives, ${\bf{F}}_{surface \rightarrow i}$ is the interaction force applied from the surface to particle $p_i$, ${\bf{F}}_{j\rightarrow i}$ is the interaction force applied from $p_j$ to $p_j$ ($i \neq j$), $\Omega_i$ is the set of neighboring particles of $p_i$, $m$ is the particle mass which is the same for all particles, and $\mu$ is a coefficient which controls the relative weights.

In the following of this section we will discuss the details of surface force calculation and how particles interact with each other.

\subsection{Surface Force}

The surface force ${\bf{F}}_{surface\rightarrow i}$ is introduced to keep particles inside of the object surface. If a particle is outside then the surface force will attract it to the object.

\begin{equation}
{{\bf{F}}_{surface \rightarrow i}} = \left\{ {\begin{array}{*{20}{c}}
0&{if{\kern 1pt} {\kern 1pt} {{\bf{n}}_i} \cdot \;{{\bf{d}}_{surface \rightarrow i}} \le 0}\\
{ - {{\bf{n}}_i} }&{if{\kern 1pt} {\kern 1pt} {{\bf{n}}_i} \cdot \;{{\bf{d}}_{surface \rightarrow i}} > 0}
\end{array}} \right.
\end{equation}

\noindent where ${\bf{d}}_{surface \rightarrow i}$ is the unit vector from to the nearest surface point to ${{p}}_i$.  ${\bf{n}}_i$ is the unit normal direction of the nearest surface point, we assume ${\bf{n}}_i$ points from inside to outside. As in Eq. \eqref{eqFisurfaceandParticle}, the weight coefficient $\mu$ is a large number to make sure the interaction forces between particles cannot push particles outside of the surface.

\subsection{Particles Interaction}

For uniformly sampling, we employed the concept that some entities repel or attract each other to avoid creating a density that is a too high or too low. We created a simple physical model by considering each particle as a balloon that is inflated gradually. The balloons will repel each other when they come in contact. In this model we need to solve two problems: (1) how to calculate the interaction force. and (2) how to inflate or deflate the balloons.

The interaction force applied from particle $p_j$ to $p_j$ is based on the simplified intermolecular forces (IMFs) model.

\begin{equation}
 h = 1/\left( {\frac{{\alpha {d_{i,j}}}}{{{r_i} + {r_j}}} + (1 - \alpha )} \right)
\end{equation}

\begin{equation}\label{eqIMFs}
{{\bf{F}}_{j \to i}} = \left( {{h^6} - {h^3}} \right){{\bf{d}}_{j \to i}}
\end{equation}

\noindent where ${{\bf{d}}_{j \rightarrow i}}$ is the unit vector from $p_j$ to $p_i$, $d_{i,j}$ is the distance between $p_i$ and $p_j$, $r_i$ and $r_j$ are the imaginary balloon radii, $\alpha$ is a coefficient. We choose $\alpha = 0.5$.

The simplified IMFs model \eqref{eqIMFs} makes interaction force zero when $d_{i,j}= r_i + r_j$. When $d_{i,j} < r_i + r_j$ , ${\bf{F}}_{j \rightarrow i} \cdot {{\bf{d}}_{j \to i}} > 0$ indicates particles will repel each other. When $d_{i,j} > r_i + r_j$ , ${\bf{F}}_{j \rightarrow i}\cdot {{\bf{d}}_{j \to i}} < 0$ indicates particles will attract each other. When $p_i$ is far away from $p_j$, their interaction force will gradually decrease to zero.


%

\begin{figure}
\vspace{0.0cm}
\centering

  \includegraphics[width=1.0\textwidth]{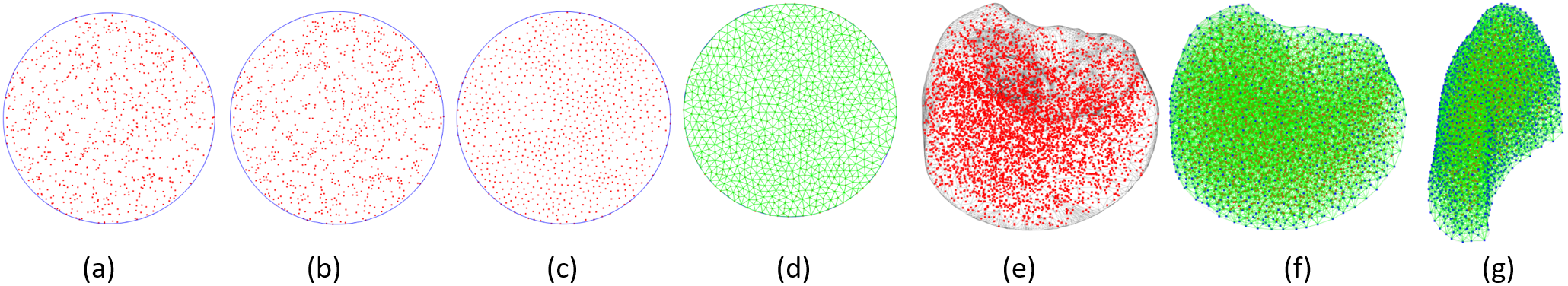}

\caption{(a)-(d) A two-dimensional example for showing our uniform sampling algorithm process. (a) the initial particles; (b)-(c) particles after 5 and 10 iterations; (d) the final particles with connections (in green line). (e)-(g) A lung lobe model sampling with 5,000 particles. (e) the initial particles; (f)-(g) the final particle with connections, the blue points are surface particles and red points are internal particles.}
\label{fig_2Dsample}
\end{figure}

In our algorithm, each balloon $p_i$ gradually dilates by imaginary inflation, but the particle interaction forces may compress the balloon.
The particle interaction force applied on $p_i$ can be divided into two parts (see \eqref{eqDivideForceintoTwoParts}). The first part on the right side of \eqref{eqDivideForceintoTwoParts} is involved in the motion of $p_i$, and the second part compresses the balloon.

\begin{equation}\label{eqDivideForceintoTwoParts}
\sum\limits_{j \in {\Omega _i}} {\left| {{{\bf{F}}_{j \rightarrow i}}} \right|}  = \left| {\sum\limits_{j \in {\Omega _i}} {{{\bf{F}}_{j \rightarrow i}}} } \right| + \left| {{{\bf{F}}_{compress,i}}} \right|
\end{equation}

It is not appropriate to make the end point of the convergence progress near $d_{i,j} = r_i + r_j$, where the interaction forces are near zero so particles take much more time to reach the goal positions. Hence, in our design, the algorithm converges at $d_{i,j} \approx 0.75 (r_i + r_j)$, where particles still repelling each other.

Considering a close-packing of equal spheres, which is a ideal uniform sampling pattern and each sphere has 12 adjacent spheres. Presuming that each balloon stops changing its radius at this ideal situation, hence:

\begin{equation}
\begin{array}{l}
\left| {{\bf{F}}_{compress,i}^{ideal}} \right| = \frac{{12}}{{{{\left( {\frac{{0.5 \times 0.75({r_i} + {r_j})}}{{{r_i} + {r_j}}} + 0.5} \right)}^6}}} - \frac{{12}}{{{{\left( {\frac{{0.5 \times 0.75({r_i} + {r_j})}}{{{r_i} + {r_j}}} + 0.5} \right)}^3}}}\; = {\rm{8}}{\rm{.8257}}
\end{array}
\end{equation}

Consequently, $r_i$ updates according to the compression force:

\begin{equation} \label{eqriupdate}
{\rm{d}}{r_i} =  - \ln (1.0 + \eta (\left| {{{\bf{F}}_{compress,i}}} \right| - \left| {{\bf{F}}_{compress,i}^{ideal}} \right|)
\end{equation}

\noindent where $\eta$ is a coefficient suggesting the hardness of each balloon.

According to \eqref{eqriupdate}, when $\left| {{{\bf{F}}_{compress,i}}} \right| < \left| {{\bf{F}}_{compress,i}^{ideal}} \right|$, the radius $r_i$ becomes larger. According to \eqref{eqIMFs}, $p_i$ will have larger interaction forces with neighboring points than before, which tends to make $\left| {{{\bf{F}}_{compress,i}}} \right|$ larger. Similarly, when $\left| {{{\bf{F}}_{compress,i}}} \right| > \left| {{\bf{F}}_{compress,i}^{ideal}} \right|$, our algorithm tends to make $\left| {{{\bf{F}}_{compress,i}}} \right|$ smaller. This forms a negative feedback system, with the result that all variables in our algorithm, including the points distribution, become stable.

For particle $p_i$, we assume it is connected to each particles in its neighboring points set

\begin{equation}
{\Omega _i} = \{ j|{d_{i,j}} < 0.6({r_i} + {r_j})\}.
\end{equation}


Fig. \ref{fig_2Dsample} shows the uniformly sampling process. Our method generates particles on both surface and inside. Those surface particles will be used as the control points in the subsequent deformation simulation and their position changes according to the change of surface.

\section{Deformation Simulation}


\begin{figure} [h]
\vspace{0.0cm}
\centering
 \subfigure[]{
  \includegraphics[width=.20\textwidth]{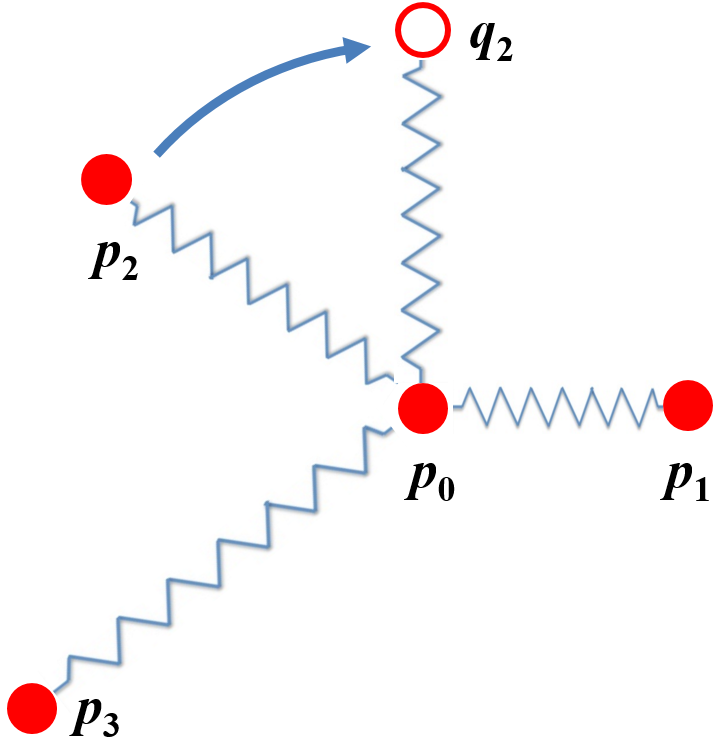}
  }
 \subfigure[]{
  \includegraphics[width=.45\textwidth]{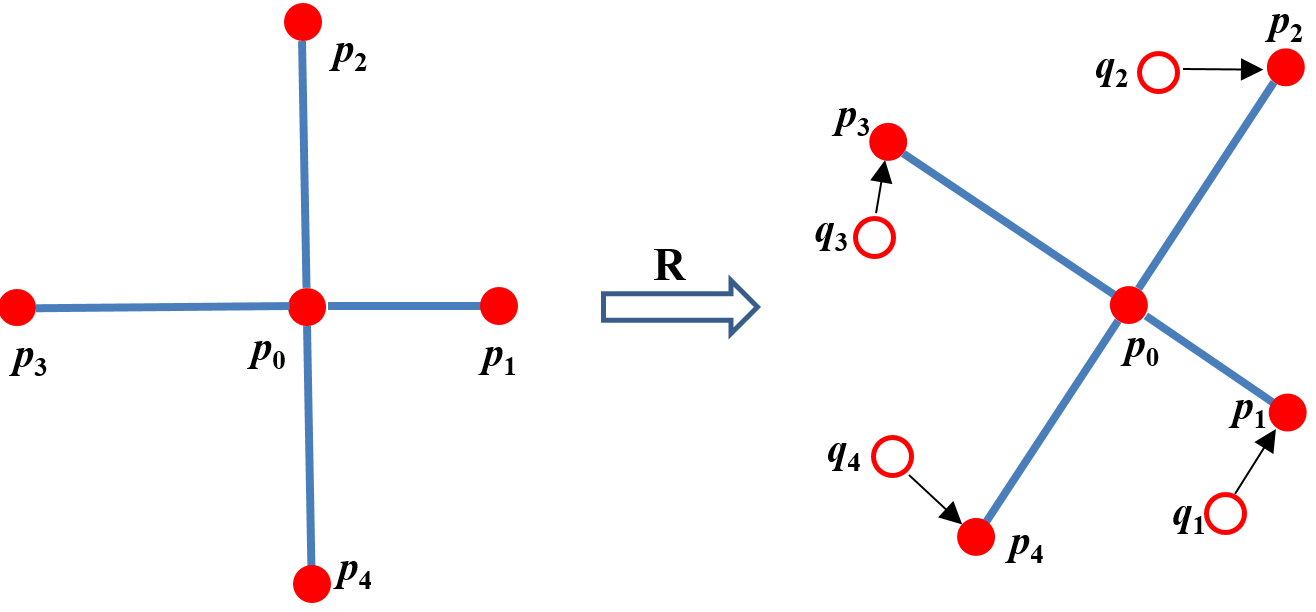}
  }
\caption{(a) If particle $p_2$ moves to $q_2$ and keeps $|p_2p_0|=|q_2p_0|$, the mass spring damper system cannot take into account this deformation and no force is generated. (b) The idea of our deformation force model. After estimating rigid rotation $\bf{R}$, the relax shape ${p_i}$ are compared with the current shape ${q_i}$ and deformation forces are generated (in black arrow). }
\label{fig_deformationforce}
\end{figure}

Our deformation simulation algorithm also utilizes particle system but with a different interaction force model. We consider the initial particles distribution and connections obtained by our uniformly sampling method as the relax state, which means there is no interaction force between particles. If surface profile varies, the surface particles, which are used as the control points, will change their positions accordingly and apply forces to internal particles. The surface particles deformation is propagated deeper into the internal structure and the final new distribution should minimize the energy of the particle system. To date, the mass spring damper system is wildly used for deformation simulation.

However, mass spring damper system has a main problem of not being able to take into account angles variation directly, as shown in Fig. \ref{fig_deformationforce}(a).
To overcome this drawback, we presented a method to calculate interaction forces by taking into account all deformation information. The idea of our method is straight forward, as shown in Fig. \ref{fig_deformationforce}(b). For a particle $p_i$ and its connected particles $p_j, j \in {\Omega _i}$, the change of their relative positions may cause interaction force. It is worth noting that the relative position variations are affected by both rigid rotation and shape deformation, we need to first get rid of the effect caused by rigid rotation.

Denoting particles at the relax state as $p$ and those at the current state as $q$, the related local shape at particle $i$ can be described in matrix form:

\begin{equation}
{{\bf{A}}_{relax,i}} = {\left[ {\begin{array}{*{20}{c}}
{{{({p_{j1}} - {p_i})}^T}}\\
 \vdots \\
{{{({p_{jn}} - {p_i})}^T}}
\end{array}} \right]_{n \times 3}},{{\bf{A}}_{current,i}} = {\left[ {\begin{array}{*{20}{c}}
{{{({q_{j1}} - {q_i})}^T}}\\
 \vdots \\
{{{({q_{jn}} - {q_i})}^T}}
\end{array}} \right]_{n \times 3}}
\end{equation}

We find a rotation matrix ${\bf{R}}_i$, which minimizes

\begin{equation}
\left\| {{{\bf{A}}_{relax,i}}{{\bf{R}}_i} - {{\bf{A}}_{current,i}}} \right\|
\end{equation}

Hence, ${\bf{R}}_i$ can be found as follows:

\begin{equation}
[{\bf{U}},{\bf{S}},{{\bf{V}}^T}] = {\mathop{\rm svd}\nolimits} ({\bf{A}}_{relax,i}^T{{\bf{A}}_{current,i}})
\end{equation}

\begin{equation}
{{\bf{R}}_i} = {\bf{U}}{{\bf{V}}^T}
\end{equation}

The deformation force applied from particle $i$ to $j$ is:

\begin{equation}
{{\bf{F}}_{i \to j}} = \sigma \left( {{\bf{R}}_i^T({p_j} - {p_i}) - ({q_j} - {q_i})} \right)
\end{equation}

\noindent where $\sigma$ is a coefficient suggesting the stiffness of the local tissue.

After calculations of all points we can get the sum of forces that a particle receives, and then we move particles according to the motion model \eqref{eqmotionmodel}.


\section{Experiments}

The algorithms proposed in this paper were implemented in C++ and VTK running on a 2.60 GHz Intel Core i7 processer. Our experiments first analyzed the algorithms runtime, and then evaluated the accuracy based on real breast tumor data.

\subsection{Runtime}

\begin{figure} [h]
\vspace{0.0cm}
\centering

  \includegraphics[width=.35\textwidth]{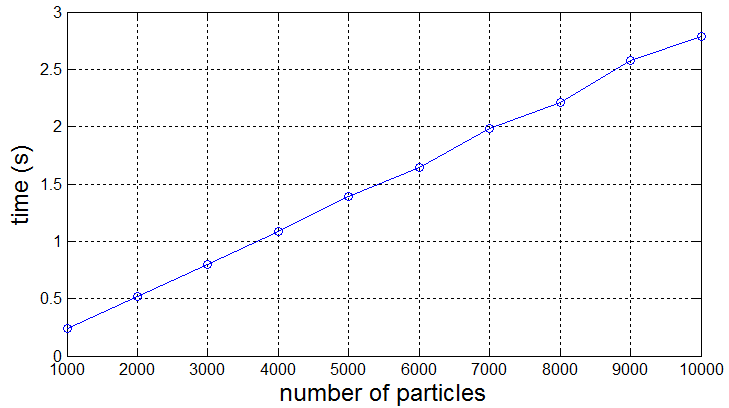}
  \includegraphics[width=.35\textwidth]{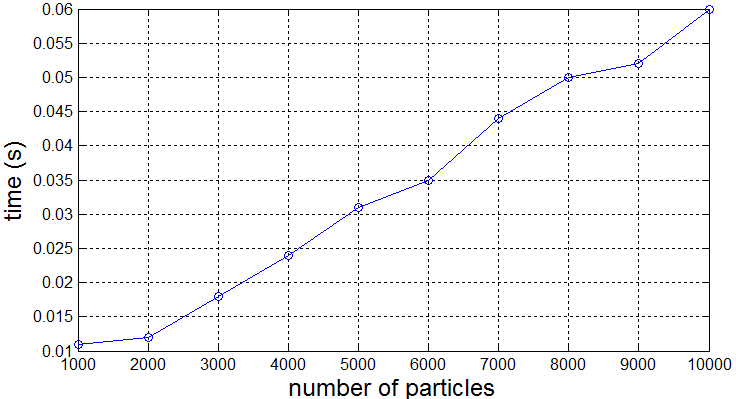}

\caption{Left: Run time of our uniformly sampling method. Right: Run time of one iteration of our deformation algorithm. Each point represents 100 trials.}
\label{fig_runtime}
\end{figure}

Both algorithms are based on particle system with linear time complexity. We evaluated their runtime with different number of particles varying from 1,000 to 10,000.

As shown in Fig. \ref{fig_runtime}(a), the uniformly sampling algorithm took less than 3 seconds for processing up to 10,000 particles. Because the uniformly sampling algorithm is only needed at the initialization stage, the real-time performance relies mostly on the deformation algorithm. As shown in Fig. \ref{fig_runtime}(b), one iteration of the deformation algorithm took an average of less than 60ms for handling up to 10,000 particles, which is very fast. It should be noted that the required number of iterations depends on how large the relative deformation is between two states, hence it is not appropriate to give an exact total runtime needed for deformation simulation. In our experiments we found that for consistently tracking a deforming object in realtime, it usually needed less than five iterations to change from one state to a new stable state. This suggests that for 10,000 particles, the average deformation runtime is around 0.3s.

\subsection{Experiments on Breast Tumor Data}


\begin{figure} [t]
\vspace{0.0cm}
\centering
\includegraphics[width=1.0\textwidth]{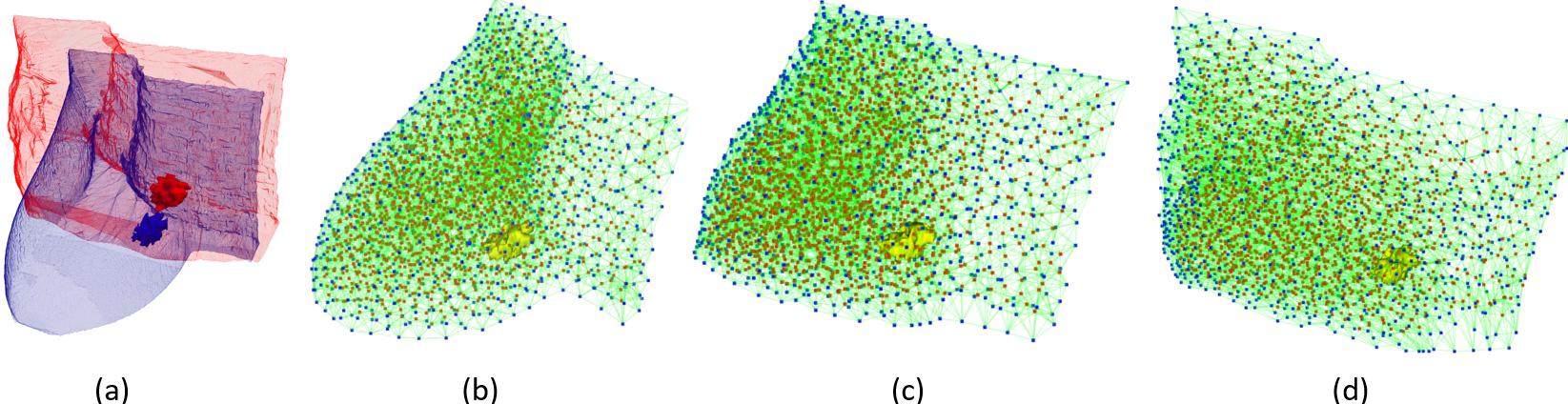}

\caption{(a) Breast and tumor segmentation results of MR image. Prone models are in blue and supine models are in red. (b)-(d) The deformation process.}
\label{fig_breastModels}
\end{figure}

\begin{figure} [h]
\vspace{0.0cm}
\centering
  \includegraphics[width=1.0\textwidth]{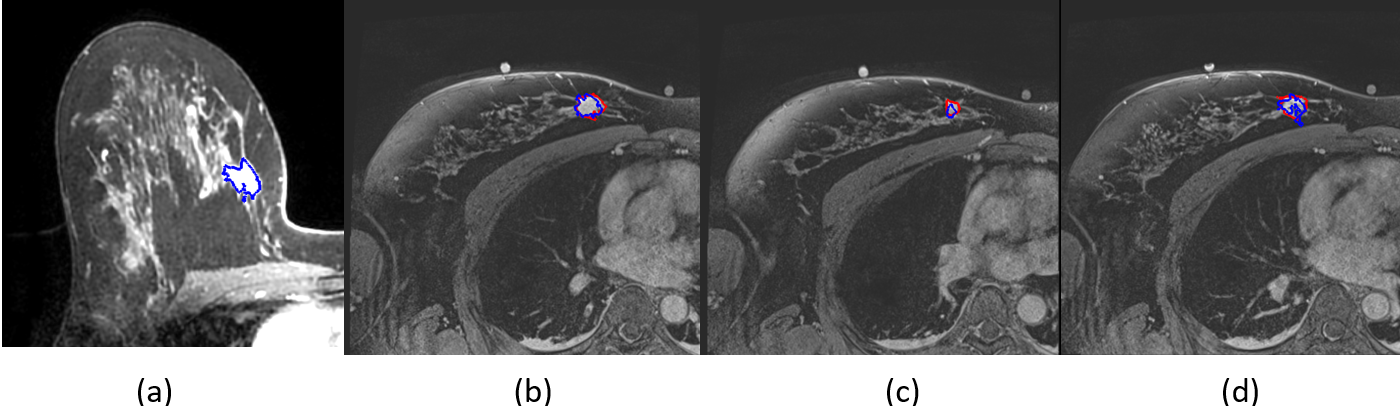}

\caption{Prone and supine breast MR imaging with segmented tumor. The blue line in (a) and red lines in (b), (c) and (d) are the tumor segmentation in prone and supine MR imaging respectively. The blue lines in (b), (c) and (d) are prone tumor deformation result.}
\label{fig_breastMRIs}
\end{figure}


Breast tumor MR images are used to evaluate the accuracy of our algorithms. We calculated the effect of breast skin deformation on the motion of the breast tumor. We segmented and modelled the chest wall, breast skin and tumor by using 3D Slicer. The breasts skin deformation were obtained by aligning intraoperative prone and supine breast MR imaging, as shown in Fig.\ref{fig_breastModels}(a). The deformation process are shown in  Fig.\ref{fig_breastModels}(b)-(d). The tumor was considered as a rigid object, whose center was represented with a particle and its orientation was adjusted according to the connected neighboring particles. The accuracy comparison results are shown in Fig. \ref{fig_breastMRIs}. Two patients' data were used and the error distance of the center of the deformed prone tumor and supine tumor were 2.8 mm and 3.5 mm respectively.


\section{Conclusion}

In this paper, we present a particle system-based method for real-time deformation of soft tissues, which achieves high accuracy with an error of several millimeters. Because the calculation of each particle is independent, the algorithms is high parallel. Future work will involve the implementation a GPU parallel computing version of the algorithms. Future work should also involve improving the particle initialization algorithm to allow generating higher density particles near region of interest to achieve high accuracy.

%
%



\begin{thebibliography}{5}
%

\bibitem{Pleijhuis}
Pleijhuis, R.G., Graafland, M., et al.,:
Obtaining adequate surgical margins in breast-conserving therapy for patients with early-stage breast cancer: current modalities and future directions. Annals of surgical oncology, 16(10), 2717-2730. (2009)

\bibitem{Conley}
Conley, Rebekah H., et al.:
Realization of a biomechanical model-assisted image guidance system for breast cancer surgery using supine MRI. International journal of computer assisted radiology and surgery 10.12, 1985-1996. (2015)

\bibitem{Le}
Le, B.H. and Deng, Z.:
Smooth skinning decomposition with rigid bones. ACM Transactions on Graphics, 31(6), 199. (2012)

\bibitem{Chao}
Chao, I., Pinkall, U., et al.:
A simple geometric model for elastic deformations. ACM transactions on graphics, 29(4), 38. (2010)


\bibitem{Macklin}
Macklin, M., Müller, et al.,:
Unified particle physics for real-time applications. ACM Transactions on Graphics, 33(4), 153. (2014)

\bibitem{MüllerSolid}
Müller, M. and Chentanez, N.:
Solid simulation with oriented particles. ACM transactions on graphics 30(92). (2011)

\bibitem{Diziol}
Diziol, R., Bender, J. , et al.:
Robust real-time deformation of incompressible surface meshes. In Proceedings of the 2011 ACM SIGGRAPH/eurographics symposium on computer animation, 237-246.(2011)

\bibitem{Shewchuk}
Shewchuk, J.R.:
Tetrahedral mesh generation by Delaunay refinement. In Proceedings of the fourteenth annual symposium on Computational geometry, 86-95. (1998)

\bibitem{SiTetGen}
Si, H.,:
TetGen, a Delaunay-based quality tetrahedral mesh generator. ACM Transactions on Mathematical Software, 41(2), 11. (2015)

\bibitem{MüllerMeshless}
Müller, M., Heidelberger, B. et al:
Meshless deformations based on shape matching. ACM transactions on graphics, 24(3), 471-478. (2005)

\bibitem{EBERLY}
Eberly, D. H.: Game Physics. Morgan Kaufmann. (2003)


\end{thebibliography}

%
%

\end{document}